\documentclass[sigconf]{acmart}

\usepackage{booktabs} % For formal tables
\usepackage{algorithm}
\usepackage[]{algpseudocode}
\usepackage{subcaption}

\usepackage{bm}

\newcommand{\MM}{\bm{M}}

\newcommand{\dd}{\bm{d}}

\newcommand{\ff}{\bm{f}}

\newcommand{\nn}{\bm{n}}
\newcommand{\xx}{\bm{x}}

\newcommand{\vv}{\bm{v}}

\newcommand{\dt}{\Delta t}

\usepackage{mathtools}

\DeclarePairedDelimiter{\floor}{\lfloor}{\rfloor}

\setcitestyle{square}

\copyrightyear{2017} 
\acmYear{2017} 
\setcopyright{acmcopyright}
\acmConference{MiG '17}{November 8--10, 2017}{Barcelona, Spain}\acmPrice{15.00}\acmDOI{10.1145/3136457.3136462}
\acmISBN{978-1-4503-5541-4/17/11}

\begin{document}
\title[Position-Based Multi-Agent Dynamics for Real-Time Crowd Simulation - Full Paper]{Position-Based Multi-Agent Dynamics \\ for Real-Time Crowd Simulation - MiG 2017}
%% \titlenote{Produces the permission block, and
%%   copyright information}

%\subtitle{paper ID: 0014}
%% \subtitlenote{The full version of the author's guide is available as
%%   \texttt{acmart.pdf} document}

\author{Tomer Weiss}
\affiliation{\institution{University of California, Los Angeles}}
\email{tweiss@cs.ucla.edu}

\author{Alan Litteneker}
\affiliation{%
 \institution{University of California, Los Angeles}
}
\email{alitteneker@cs.ucla.edu}

\author{Chenfanfu Jiang}
\affiliation{%
  \institution{University of Pennsylvania}
}
\email{cffjiang@seas.upenn.edu}

\author{Demetri Terzopoulos}
\affiliation{%
  \institution{University of California, Los Angeles}
}
\email{dt@cs.ucla.edu}

% The default list of authors is too long for headers}
\renewcommand{\shortauthors}{T.~Weiss, A.~Litteneker, C.~Jiang, and D.~Terzopoulos}

\begin{abstract}
Exploiting the efficiency and stability of Position-Based Dynamics
(PBD), we introduce a novel crowd simulation method that runs at
interactive rates for hundreds of thousands of agents. Our method
enables the detailed modeling of per-agent behavior in a Lagrangian
formulation. We model short-range and long-range collision avoidance
to simulate both sparse and dense crowds. On the particles
representing agents, we formulate a set of positional constraints that
can be readily integrated into a standard PBD solver. We augment the
tentative particle motions with planning velocities to determine the
preferred velocities of agents, and project the positions onto the
constraint manifold to eliminate colliding configurations. The local
short-range interaction is represented with collision and frictional
contact between agents, as in the discrete simulation of granular
materials. We incorporate a cohesion model for modeling collective
behaviors and propose a new constraint for dealing with potential
future collisions. Our new method is suitable for use in interactive
games.
\end{abstract}

%
% The code below should be generated by the tool at
% http://dl.acm.org/ccs.cfm
% Please copy and paste the code instead of the example below. 
%



\begin{teaserfigure}
\begin{center}
\begin{subfigure}{0.49\textwidth}
\includegraphics[width=\textwidth]{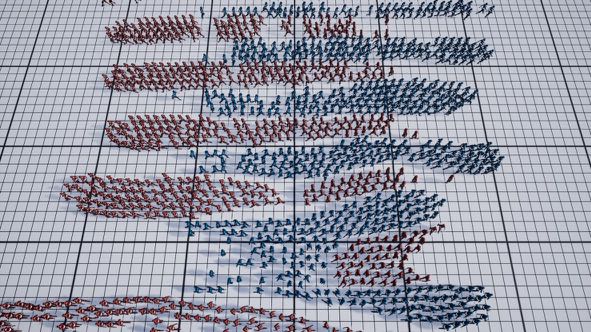} 
\label{fig:teaser1}
\end{subfigure} \hfill
\begin{subfigure}{0.49\textwidth}
\includegraphics[width=\textwidth]{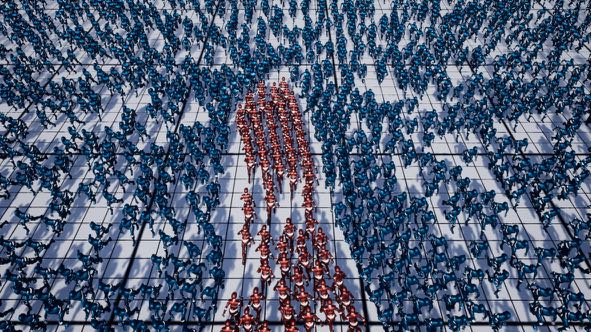}
\label{fig:teaser2}
\end{subfigure}
\end{center}
  \vskip -10pt
  \caption{Our PBD-based crowd simulation method animates both sparse
  and dense groups of agents at interactive rates.
  \label{fig:all_teaser}}
  \vskip 20pt
\end{teaserfigure}

\maketitle

\section{Introduction}

Crowd simulation is ubiquitous in visual effects, animations, and
games. Efficiently simulating the motions of numerous agents with
realistic interactions among them has been a major focus of research
in recent decades \cite{thalmann2007crowd}. Among various modeling
considerations, collision avoidance remains challenging and time
consuming. Collision avoidance algorithms can be classified into
discrete and continuum approaches \cite{golas:2013:hybrid}. Continuum
approaches, such as the technique proposed by Narain et
al.~\shortcite{narain:2009:aggregate}, have proven efficient for
large-scale dense crowds, but are less suitable for sparse crowds.
Force-based discrete approaches, such as the recently proposed
power-law model \cite{karamouzas:2014:powerlawPRL}, are well suited
for sparse crowds, but can be computationally expensive and may
require smaller time steps due to explicit time integration.

In this paper, we use Position-Based Dynamics (PBD)
\cite{muller:2007:pbd}, as an alternative discrete algorithm for
simulating both dense and sparse crowds. While more carefully designed
models, such as the social force model \cite{helbing1995social} and
the power law model \cite{karamouzas:2014:powerlawPRL}, can yield some
realistic crowd behaviors, they occasionally require elaborate
numerical treatments to remain stable and robust. Given the success of
PBD in simulating various solid and fluid materials in real-time
physics, our work further extends the idea to crowd simulation.

Our objective is therefore to offer a numerical framework for crowd
simulation that is robust, stable, and easy to implement, ideally for
use in interactive games. Due to the flexibility of PBD in defining
positional constraints among particles, our proposed framework
provides a new platform for artistic design and control of agent
behaviors in crowd modeling and animation. Furthermore, we adopt the
PBD approach since it is an unconditionally stable implicit scheme.
Even though it may not always converge to the solution manifold, a
nonlinear Gauss-Seidel-like constraint projection enables the
algorithm to produce satisfactory results with modest computational
cost suitable for real-time applications. Additionally, the resulting
solution scheme is easy to implement and does not require any linear
solves.

\subsection{Contributions}

This paper, which extends \cite{Weiss2017position}, makes the
following contributions:
\begin{itemize}
\item We show how crowds can be simulated within the PBD framework by
augmenting it with non-passive agent-based planning velocity.
\item We adopt position-based frictional contact constraints of
granular materials to model local collision avoidance among nearby
agents. An XSPH viscosity term is also added to approximate coherent
and collective group behavior.
\item We develop a novel long-range collision avoidance constraint to
deal with anticipatory collisions. Our model permits the natural
development of agent groups.
\item We demonstrate multi-species crowd coupling by supporting
spatially varying Lagrangian physical properties.
\end{itemize}

\subsection{Overview}

The remainder of the paper is organized as follows:
Section~\ref{sec:related} surveys relevant prior work on crowd
simulation and PBD. Section~\ref{sec:overview} overviews our
algorithmic approach. Section~\ref{sec:method} discusses algorithmic
details and detailed constraint design. We present our simulation
results in Section~\ref{sec:result}. Section~\ref{sec:discuss}
concludes the paper with a discussion of our method's limitations and
future work.

\section{Related Work}
\label{sec:related}

Position-Based Dynamics (PBD) was first introduced by M\"{u}ller et
al.~\shortcite{muller:2007:pbd} for the fast simulation of deformable
objects through the Gauss-Seidel projection of positional constraints.
Since then, PBD and Nucleus, a closely-related constraint solver by
Stam~\shortcite{stam2009nucleus}, have become popular in physics-based
animation for their simplicity and robustness. Macklin et
al.~\shortcite{macklin:2014:PBD} presented a unified PBD solver for
various natural phenomena. XPBD was proposed recently to eliminate the
iteration count and time step dependence of PBD
\cite{Macklin:2016:XPS:2994258.2994272}. Even though PBD traditionally
defines geometric constraints among particles, it can also approximate
force responses from continuum mechanics. Bender et
al.~\shortcite{bender2014position} formulated continuum energies as
PBD constraints.
%% M\"{u}ller et al~\shortcite{Muller:2015:SBD} defined
%% constraints based on element strain measures.
The close relationship between PBD and popular
continuum-mechanics-based discretization was further explored in the
recent work on optimization-based methods for real-time animation
\cite{liu:2013:fsm,bouaziz:2014:pdf,Wang:2015:CSA,Narain2016ADMM}. A
more complete survey of PBD is provided by Bender et
al.~\shortcite{bender2014survey}.

Efficient, natural, and stable collision avoidance in sparse and dense
distributions of agents remains a very active area of research in
crowd simulation. In multi-agent simulations, it is essential to
capture both individual local behaviors and aggregate collective
behaviors. Continuum approaches---such as `Continuum Crowds'
\cite{treuille2006continuum}, a crowd model that uses continuum
dynamics to simulate pedestrian flow---are particularly suitable for
dense crowds and complex environments
\cite{jiang2010continuumcomplex}.
%% Continuum models are also used in traffic simulation
%% \cite{sewall2010continuum}. As in pedestrian simulations, hybrid
%% methods that combine agent-based behaviors and continuum dynamics
%% yield more realistic results, as in \cite{sewall2011traffic}.
Unfortunately, the traditional regime of pure continuum models tends
to smooth out local agent behaviors, motivating research on hybrid
methods. For example, Narain et al.~\shortcite{narain:2009:aggregate}
simulated dense crowds with a hybrid, Eulerian-Lagrangian
particle-in-cell approach and the unilateral incompressibility
constraint (UIC), which has proven to be an effective assumption for
crowds.
%% but also for granular flow \cite{ivan2011sand,narain:2010:sand} and
%% liquids \cite{gerszewski:2013:splashing}.
Subsequently, frictional forces were taken into account in modeling
crowd turbulence \cite{helbing2007dynamics, golas:2014:cmc}, which is
essential in extra high-density scenarios. This also inspired us to
treat dense agent collisions with a frictional contact model similar
to PBD dry sand simulation in \cite{macklin:2014:PBD}. To robustly
model multiple densities, Golas et al.~\shortcite{golas:2013:hybrid}
proposed a hybrid scheme for simulating high-density and low-density
crowds with seamless transitions.

Other techniques for collision avoidance have been proposed. Many
researchers adopted force-based models
\cite{Reynolds:1987,reynolds1999steering,helbing2000simulating}. An
interaction energy between pedestrians was modeled with a power law in
recent and concurrent work by Karamouzas et
al.~\shortcite{karamouzas:2014:powerlawPRL,karamouzas17}. As an
alternative to forces, the reciprocal velocity obstacle was proposed
in robotics for multi-agent navigation \cite{van2008reciprocal}. Guy
et al.~\shortcite{guy2009clearpath} extended velocity obstacles and
used a parallel optimization framework for collision avoidance. Ren et
al.~\shortcite{ren2016group} augment velocity obstacles with velocity
connections to keep agents moving together, thus allowing more
coherent behaviors. He et al.~\shortcite{he:2016:group} simulated
dynamic group behaviors based on the least effort principle. Guy et
al.~\shortcite{guy2010ple} simulated large-scale crowds by
optimization based on the Principle of Least Effort.
%% Yeh at el.~\shortcite{yeh2008composite} introduced composite agents
%% for complex agent interactions.
Bruneau and Pettr{\'e}~\shortcite{bruneau2015midterm} presented a
mid-term planning system to fill in the gap between long-term planning
and short-term collision avoidance.
%% In the work of Kim et al.~\shortcite{kim2013physical}, multi-agent
%% simulation and physical interaction with obstacles were nicely
%% combined to generate interesting new behaviors.

\begin{algorithm}
\caption{Position-Based Crowd simulation loop}\label{alg:loop}
\begin{algorithmic}[1]
\ForAll{agent i} \Comment{\S \ref{sec:planning}}
\State calculate $\vv_i^p$ from a velocity planner 
\State calculate a blending velocity $\vv_i^b$ from $\vv_i^p$ and $\vv_i^n$ 
\State $\xx_i^* \gets \xx_i^n + \dt \vv_i^b$
\EndFor
\ForAll{agent i}
\State find neighboring agents $S_i = \{s_{i1}, s_{i2},...,s_{im}\}$
\EndFor
\While{iteration count$<$ max stability iterations}
\ForAll{agent i}
\State compute position correction $\Delta \xx_i$ \Comment{\S \ref{sec:contact}}
\State $\xx_i^n \gets \xx_i^n + \Delta \xx_i$
\State $\xx_i^* \gets \xx_i^* + \Delta \xx_i$
\EndFor
\EndWhile
\While{iteration count$<$ max iterations}
\ForAll{agent i}
\State compute position correction $\Delta \xx_i$ \Comment{\S \ref{sec:contact},
  \S \ref{sec:long}, \S \ref{sec:slide}}
\State $\xx_i^* \gets \xx_i^* + \Delta \xx_i$
\EndFor
\EndWhile
\ForAll{agent i}
\State $\vv_i^{n+1} \gets (\xx_i^* - \xx_i^n)/\dt$
\State Add XSPH viscosity to $\vv_i^{n+1}$ \Comment{\S \ref{sec:cohesion}}
\State Clamp $\vv_i^{n+1}$ \Comment{\S \ref{sec:limiting}}
\State $\xx_i^{n+1} \gets \xx_i^*$ 
\EndFor
\end{algorithmic}
\end{algorithm}

\section{Algorithm Overview}
\label{sec:overview}

Our simulation loop per time step is similar to that for PBD, with
several modifications. We outline our procedure in
Algorithm~\ref{alg:loop} and highlight the different steps.

Assume that we have $N$ agents. Each agent $i$, where $i=1,2,\dots,N$,
is represented with a fixed-sized particle with position $\xx_i \in
\mathbb{R}^2$ and velocity $\vv_i \in \mathbb{R}^2$. For multi-species
considerations, we treat each particle as a circle with radius $r_i$
and mass $m_i$. When we are stepping from time $n$ to time $n+1$ in a
traditional PBD simulation loop for passive physical simulations, a
forward Euler position prediction is first performed as $\xx^*_i =
\xx_i^n + \dt (\vv_i^n + \dt \ff_\mathrm{ext}(\xx_i^n))$, where
$\ff_\mathrm{ext}$ represents external forces such as gravity. In
position-based crowds, $\xx^*_i$ needs to be computed differently to
take into account the velocity planning of each agent. In particular,
we compute $\xx_i^*$ based on a blending scheme between a preferred
velocity and the current velocity $\vv_i^n$ (see
Section~\ref{sec:planning}). It is apparent that the predicted
$\xx^*_i$ for a particle completely ignores the existence of any other
particles and just passively advects in the velocity field. To resolve
this, PBD defines constraint functions on the desired location of the
particles. Both equality and inequality constraints are supported, and
they can be expressed as $C_k(\xx_1,\xx_2,...,\xx_N) = 0$ and
$C_k(\xx_1,\xx_2,...,\xx_N)\ge 0$ respectively. Hence, the task is to
search for a correction $\Delta \xx_i$ such that $\xx_i^{n+1} =
\xx^*_i + \Delta \xx_i$ satisfies the constraints. Once the new
positions are computed, agent velocities can be updated as
$\vv_i^{n+1} = (\xx_i^{n+1}-\xx_i^n)/\dt$. This update guarantees
stable agent velocities as long as the constraint projection is
stable.

\section{Method}
\label{sec:method}

Our position-based formulation includes several modifications to the
standard PBD scheme as well as additional constraints for short-range
and long-range collision avoidance between agents.

\subsection{Velocity Blending}
\label{sec:planning}

Agent level roadmap velocity planning describes high-level agent
behaviors. Local behavior may be influenced by factors such as social
or cognitive goals, while global behavior may be specified by a
particular walking path.
%% \cite{Shao:2007,Yu:2007:DNF:1272690.1272707}.
We note that the roadmap planning is an orthogonal component to our
constraint based scheme.

In the physics-based simulation of solids and fluids, particles
generally retain their existing velocities. In particular, as
demonstrated in \cite{bouaziz:2014:pdf}, the implicit Euler time
integration of a physical system can be formulated as an minimization
problem that balances the `momentum potential' $\|
\MM^{1/2}(\xx-(\xx^n+\dt \vv^n))\|_F^2/2\dt^2$ and other potential
energies, where $\MM$ is the mass matrix. In a multi-agent crowd
simulation, it is similarly more desirable to include the inertia
effect before predicting an agent's desired velocity. Denoting the
preferred velocity given the planner with $\vv_i^p$, we calculate the
agent velocity $\vv_i^b$ as a linear blending between $\vv_i^p$ and
the current velocity $\vv_i^n$, as follows:
\begin{equation}
  \vv_i^b = (1-\alpha) \vv_i^n + \alpha \vv_i^p,\label{eqn:vib}
\end{equation}
where $\alpha \in [0,1]$. We set $\alpha=0.0385$ in all our
simulations. A more adaptive choice, such as the density-based
blending factor as in \cite{narain:2009:aggregate}, can also be used
in our framework.

\subsection{Frictional Contact}
\label{sec:contact}

We model local particle contacts with an inequality distance
constraint as in standard position-based methods:
\begin{equation}
C(\xx_i,\xx_j) = \|\xx_i - \xx_j\| - (r_i+r_j) \ge 0, \label{eq:traditional} 
\end{equation}
where $r_i$ and $r_j$ are the radii of agents $i$ and $j$. To model
frictional behavior between neighboring agents, we further adopt
kinematic frictions as described in \cite{macklin:2014:PBD}.

\subsection{Cohesion}
\label{sec:cohesion}

To encourage more coherent agent motions, we add the artificial XSPH
viscosity \cite{Schechter:2012,Macklin:2013:PBF} to the updated agent
velocities. Specifically,
\begin{equation}
\vv_i \gets \vv_i + c \sum_j (\vv_i-\vv_j) W(\xx_i - \xx_j,h),
\end{equation}
where $W(\bm{r},h)$ is the Poly6 kernel for SPH
\cite{Macklin:2013:PBF}. For our simulations, with particles with
radius $1$, we use $h=7$ and $c=217$.

\begin{figure*}
\includegraphics[width=\textwidth]{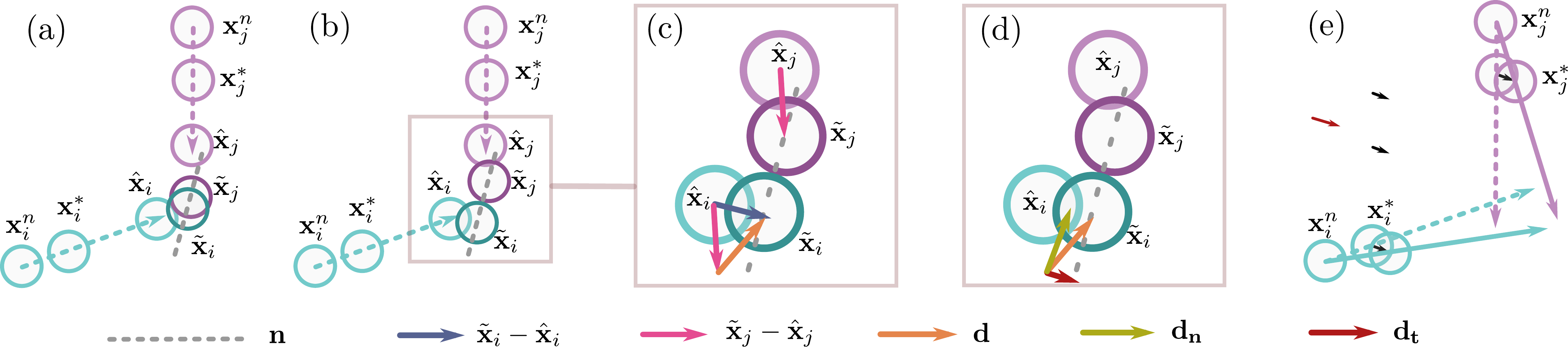}
\caption{Avoidance model for predictive collision avoidance. (a)
Starting with particles in current positions $\xx_i^n$ and $\xx_j^n$,
PBD estimates their positions $\xx_i^*$ and $\xx_j^*$ at the next time
step. To further predict behaviors in the future, we estimate a
discrete time to collision $\hat{\tau}$ using their trajectories. This
results in $\hat{\xx}_{i,j} = \xx_{i,j}^n + \hat{\uptau} \vv_{i,j}$.
When further advanced in time by $\dt$, particles collide at
$\tilde{\xx}_i$ and $\tilde{\xx}_j$. (b) Projecting these collision
constraints resolves the collision between $\tilde{\xx}_i$ and
$\tilde{\xx}_j$. (c) We compute the relative displacement $\dd$ from
time $\hat{\uptau}$ to $\tilde{\uptau}$. (d) $\dd$ is decomposed into
contact normal ($\dd_{\nn}$) and tangential ($\dd_{\bm{t}}$)
components. (e) The tangential contribution of the relative
displacement is distributed to $\xx_i^*$ and $\xx_j^*$, which results
in an avoidance resolution of future contacts.
  \label{sliding}}
\end{figure*}

\subsection{Long Range Collision}
\label{sec:long}

Karamouzas et al.~\shortcite{karamouzas:2014:powerlawPRL} describe an
explicit force-based scheme for modeling crowds. We design a similar
scheme as a position-based constraint. As in their power law setting,
the leading term is the time to collision $\uptau$, defined as the
time when two disks representing particles $i$ and $j$ touch each
other in the future. As in \cite{karamouzas:2014:powerlawPRL}, it can
be shown that
\begin{equation}
\tau = \frac{b-\sqrt{b^2-ac}}{a},\label{eqn:tau}
\end{equation}
where
\begin{align}
a&=\frac{1}{\dt^2}\|\xx_i^*-\xx_j^*\|^2, \\
b&=-\frac{1}{\dt}(\xx_i-\xx_j)\cdot(\xx_i^*-\xx_j^*), \\
c&=\|\xx_i-\xx_j\|^2 - (r_i+r_j)^2.
\end{align}

No potential energies associated with forces are required in our
framework. To facilitate collision-free states in the future, we
directly apply a collision-free constraint on future positions. Recall
that in our simulation loop, the predicted position of particles $i$
and $j$ in the next time step are
\begin{equation}
  \xx_{i,j}^* = \xx_{i,j}^n + \dt \vv_{i,j},
\end{equation}
where $\vv_{i,j}^b$ is defined in (\ref{eqn:vib}) , and we use the
$i,j$ subscripts to denote that the above is defined exclusively in
the context of $i$ or $j$.

We estimate a future collision state between $i$ and $j$ using
$\uptau$. We first compute the exact time to collision using
(\ref{eqn:tau}). Valid cases are those with $\uptau>0$ and $\uptau <
\uptau_{0}$, where $\uptau_{0}$ is a fixed constant. We used
$\uptau_{0}=20$ in all our experiments. After pruning out invalid
cases, we process the remaining colliding pairs in parallel (Section~\ref{sec:params}).  
We define $\hat{\uptau}
= \dt*\floor*{\uptau/\dt}$, where $\floor*{\cdot}$ denotes the floor
operator. This is simply clamping $\uptau$ to find a discrete time
spot slightly before the predicted contact. With $\hat{\uptau}$, we
have
\begin{equation}
  \hat{\xx}_{i,j} = \xx_{i,j}^n + \hat{\uptau} \vv_{i,j}.
\end{equation}
Note $\hat{\xx}_{i,j}$ are similar to $\xx_{i,j}^n$ in the traditional collision
constraint case (\ref{eq:traditional}) and are still in a collision free state. Stepping forward will
cause the actual penetration. We define the colliding positions with
\begin{equation}
  \tilde{\xx_{i,j}} = \xx_{i,j}^n + \tilde{\uptau} \vv_{i,j},
\end{equation}
where $\tilde{\uptau}=\dt+\hat{\uptau}$. We enforce a collision free
constraint on $\tilde{\xx_i}$ and $\tilde{\xx_j}$. Note that
$\tilde{\xx}_{i,j}$ is a function of $\xx_{i,j}^*$; therefore, it is
still essentially a constraint on $\xx_{i,j}^*$. Due to its
anticipatory nature, high stiffness on this constraint is not
necessary. To prevent over-stiff behaviors, instead of using the
overlap between the predicted particle locations, we define the
stiffness to be $k \exp(-{\hat{\uptau}}^2 / {\uptau_0} )$, where $k$ is
a user-specified constant.

\subsection{Avoidance Model}
\label{sec:slide}

We further present a novel avoidance model for crowd collision. 
The long-range collision constraint from Section~\ref{sec:long}
will cause agents to slow down due to motion along the contact normal
from the collision resolve, which is often not desirable in dense
scenarios (Fig.~\ref{fig:all_teaser}). However, we observe that the
tangential component of that collision response is often desired,
effectively causing the agents to simply slide in response to the
predicted collision. Hence, we preserve only the tangential movement
in such collisions. We calculate the total relative displacement as
 \begin{equation}
   \dd = (\tilde{\xx}_i -\hat{\xx}_i)- (\tilde{\xx}_j -\hat{\xx}_j),
 \end{equation}
which can be decomposed into contact normal and tangential components
as follows:
 \begin{align}
   \dd_{\nn} &= (\dd\cdot\nn)\nn, \\
   \dd_{\mathbf{t}} &= \dd - \dd_{\nn},
 \end{align}
where
$\nn=(\tilde{\xx}_i-\tilde{\xx}_j)/{\|\tilde{\xx}_i-\tilde{\xx}_j\|}$
is the contact normal. To this end, we preserve only the tangential
component in the positional correction to $\xx_{i,j}^*$. This provides
an avoidance behavior and prevents agents from being pushed back in a
dense flow. Fig.~\ref{sliding} illustrates this process.

\subsection{Maximum Speed and Acceleration Limiting}
\label{sec:limiting}

After the constraint solve, we further clamp the maximum speed and acceleration 
of the agents to better approximate real human capabilities.

\subsection{Walls and Obstacles}
\label{sec:walls}

Agents can interact with walls and other static obstacles in the
environment. We prevent agents locomoting into walls and other static
obstacles by a traditional collision response (\ref{eq:traditional}),
between the agent's predicted position and the nearest point on the
obstacle. The obstacle's collision point is assigned infinite mass.

\begin{figure}
\includegraphics[width=\columnwidth]{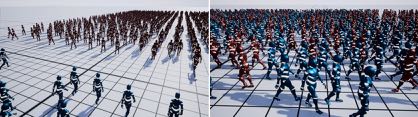}
\caption{Two groups of agents passing through each other using our
long range collision avoidance.
\label{single_z}}
\end{figure}

\begin{figure*}
\includegraphics[width=\textwidth]{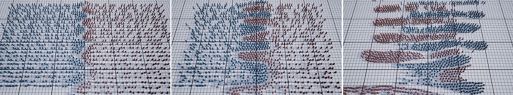}
\caption{Groups passing each other using the avoidance model. Left:
Groups of agents organize into a boundary front in preparation for
collision avoidance. Middle: Agents huddle together in noticeable
thick lanes. Right: Agents successfully pass each
other.
\label{double_sparse}}
\end{figure*}

\section{Experiments and Results}
\label{sec:result}

\subsection{Setup and Parameter settings}\label{sec:params}

We implemented our framework in CUDA, using an NVIDIA GeForce GT 750M.
We set $\dt=1/48$ sec for all the experiments (2 substeps per frame).
We solve each constraint group in parallel, employing a Jacobi solver,
with a delta averaging coefficient of 1.2. To find neighboring agents,
we use two hash-grids, for short and long range collisions. This is
more efficient than using one grid for both, since the long range grid
covers a bigger collision radius. Each grid is constructed efficiently
and in parallel. See \cite{green2008cuda,macklin:2014:PBD} for
additional details.

In our simulations, we use 1 stability iteration to resolve contact
constraints possibly remaining from the previous time step, and 6
iterations in the constraint solve loop. Additional iterations can
increase stability and smoothness, but at increased computational
cost.

For agent rendering and locomotion synthesis, we used Unreal Engine
4.15. For smooth locomotion, we clamped the agent's skeletal
positional acceleration and rotational velocity. Additionally, we
applied a uniform motion scaling of about 30. We rendered the motion
at about 5 times the simulation rate.

We demonstrated the robustness of our position-based framework in a
variety of scenarios. To simplify the experiment setup and unless
otherwise stated, we modeled all agents using a disk with radius 0.5,
and use the same width for our humanoid agents in the rendering stage.
For smoother motion, we allow an expansion of the agent's disk radius
by 5\% during collision checks. For each benchmark, we used a simple
preferred velocity planner, where the preferred velocity of each agent
points to the closest user-scripted goal. We also slightly varied the
preferred velocity of each agent around a mean of 1.4, to achieve a
more realistic simulation. Table~\ref{table:timings} presents timing
information.

\begin{table}
\centering
\fontsize{.85em}{.7em}\selectfont
%\begin{tabular}{@{}l@{}l@{\,\,\,\,}l@{\,\,\,\,}l@{\,\,\,\,}}
\begin{tabular}{llllll}
\toprule
&    \textbf{\# agents} & \textbf{LR} & \textbf{A} & \textbf{ms/frame}   \\ \midrule
Sparse passing  &  1,600  &  On   &  -  & 11.27     \\ %%%%
Sparse passing  &  1,600  &  -    & On  & 11.61      \\ %%%%
Dense, low count  &  1,600  &  On   & -   & 12.03     \\ %%%%
Dense, low count  & 1,600   &  -    & On  & 11.34     \\ %%%%
Dense, high count  &  10,032  &  On   & -   & 14.06   \\ %%%%
Dense, high count  & 10,032   &  -    & On  & 13.63   \\ %%%%
Bears and Rabbits  & 1,152 &  -  & On   & 11.86      \\ %%%%
Dense Ellipsoid   &  1,920  &  -  & On   & 10.06     \\ %%%%
Proximal Behavior   &  50  &  On  & -   & 10.12      \\ %%%%
Proximal Behavior   &  50  &  -   & On  & 10.13      \\ %%%%
Target Locomotion   &  192 &  On  & -   & 10.42       \\ %%%%
Bottleneck &  480   &  -  & -   & 11.99       \\ %%%%
Bottleneck &  3,600   &  -  & -   & 17.76    \\ %%%%
Bottleneck &  100,048   &  -  & -   & 43.66 \\ %%%%
\bottomrule
\end{tabular}
\caption{Timings. LR: long range collision constraint; A: avoidance
model constraint. All experiments use $\dt=1/48$, with 6 iterations
per time step. These timings do not include rendering times.
\label{table:timings}}
\end{table}

\subsection{Benchmarks and Analysis}

\subsubsection{Sparse Passing (Low Count, Long Range Collision): }
\label{exp:sparse_long_low}
We experimented with two groups of agents locomoting in opposite
directions (Fig.~\ref{single_z}). The agents in each group are
positioned in a loose grid formation with an initial separation
distance. To avoid collisions, the agents use the constraint of
Section~\ref{sec:long}. In this scenario, the agents organize into
narrow lanes, and pass each other easily.

\subsubsection{Sparse Passing (Low Count, Avoidance): }
\label{exp:sparse_slide_low}
This scenario is identical to \ref{exp:sparse_long_low}, but the
agents employ the constraint of Section~\ref{sec:slide} to avoid
collisions. In this scenario, the agents form thicker lanes
(Fig.~\ref{double_sparse}), which accumulate into different groups.

\begin{figure}
\includegraphics[width=\columnwidth]{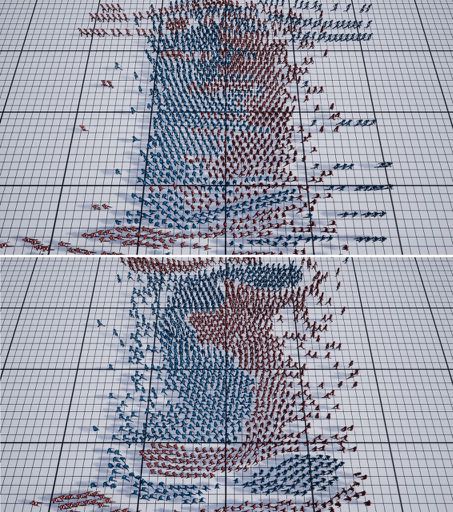}
\caption{High density agent simulation. Top: Long range collision.
Bottom: Avoidance model.
\label{single_dense}}
\end{figure}

\subsubsection{Dense Passing (Low Count, Long Range Collision): }
\label{exp:dense_long_low}
A total of 1,600 agents are split into two groups, with a separating
distance of 2.5 (Fig.~\ref{single_dense}). We used a higher and denser
crowd of agents. To avoid collision, the agents employ the constraint
of Section~\ref{sec:slide}. Because of the dense agent setting, the
two agent groups do not easily pass each other, and some bottleneck
groups are formed. Eventually, the agents pass, avoiding unrealistic
collisions.

\subsubsection{Dense Passing (Low Count, Avoidance): }
\label{exp:dense_slide_low}
This experimental setup is identical to \ref{exp:dense_long_low}. To
avoid collision, the agents employ the constraint of
Section~\ref{sec:slide}. In this scenario, the agents form thicker
lanes, which form into different groups (Fig.~\ref{single_dense}).

\begin{figure*}
\includegraphics[width=\textwidth]{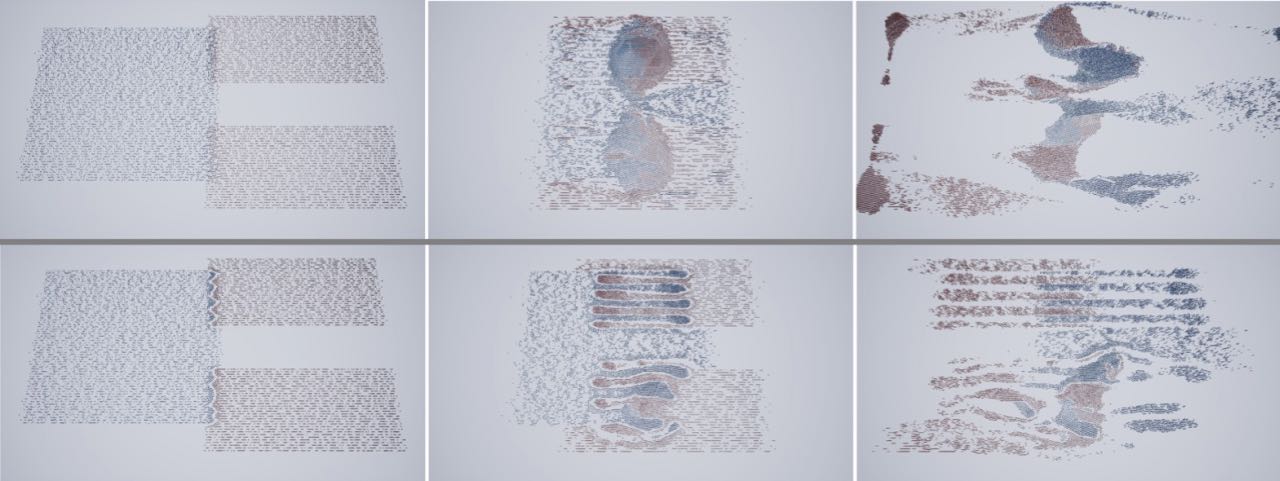}
\caption{High density and high agent count. Top Row: Agent groups
avoid each other using Long Range Collision. Bottom Row: Using the
Avoidance model.
\label{double_dense_high_count}}
\end{figure*}

\subsubsection{Dense Passing (High Count, Long Range Collision): }
\label{exp:pass_long_high}
A total of 10,032 agents are split into two groups
(Fig.~\ref{double_dense_high_count}) with a separating distance of
3.5. This experiment setup is identical to \ref{exp:pass_long_high}.

\subsubsection{Dense Passing (High Count, Avoidance): }
\label{exp:pass_slide_high}
This experiment setup is identical to \ref{exp:pass_long_high}. To
avoid collision, the agents employ the constraint of
Section~\ref{sec:slide}. In this scenario, the agents form thicker
lanes, which form into different groups.

\begin{figure}
\includegraphics[width=\columnwidth]{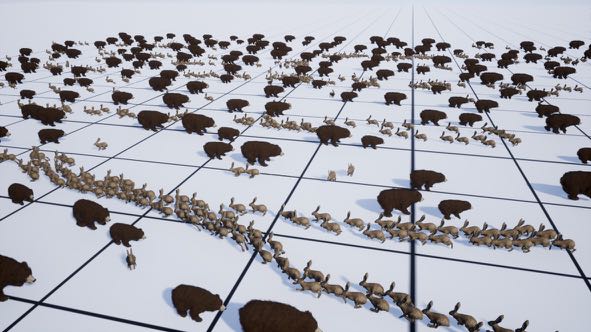}
\caption{A group of smaller agents (rabbits) passing through a group
of larger ones (bears).
\label{single_bear}}
\end{figure}

\begin{figure}
\includegraphics[width=\columnwidth]{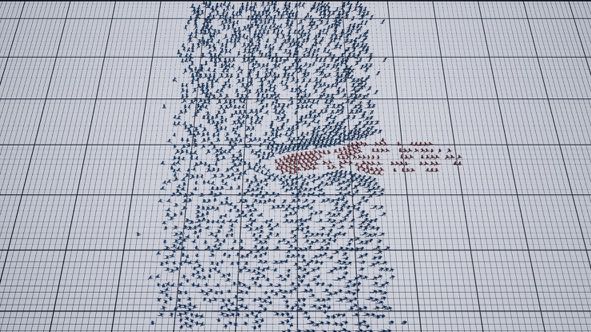}
\caption{A small ellipsoid shaped group passing through a larger
group.
\label{single_boat}}
\end{figure}

\subsubsection{Bears and Rabbits: }
\label{exp:bears}
In this experiment, we showcased how a Lagrangian PBD scheme may be
employed to model agents of different sizes (Fig.~\ref{single_bear}).
We modeled a group of rabbits passing through a group of bears,
totaling 1,152 agents. The rabbits had size 1.0, while the bears had a
size ranging from 2.5 to 4.0. To simulate that bears are less prone to
change their path than rabbits, we assigned the bears a mass that is
approximately 30 times greater than that of the rabbits.

\begin{figure*}
\includegraphics[width=\textwidth]{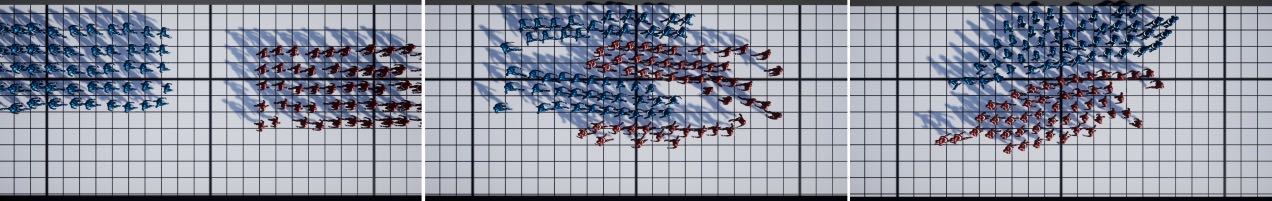}
\caption{Proxemic group behavior. Left: Initial state. Center: Agents
avoid each other using the long-range collision model, while creating
lanes. Right: Agents avoid each other using the avoidance model.
\label{double_proxemic}}
\end{figure*}

\begin{figure*}
\includegraphics[width=\textwidth]{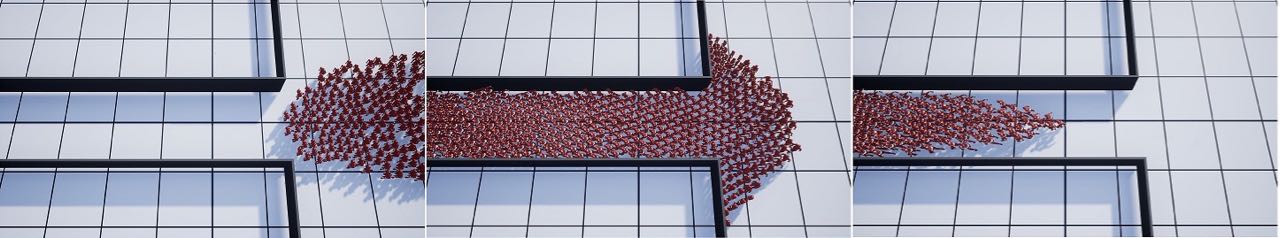}
\caption{A group of agents passing through a narrow corridor. Left:
Agents huddle on approaching corridor's entrance. Middle: A
semi-circular arch forms as agents enter a narrow corridor. Right:
Agents successfully exit.
\label{arching}}
\end{figure*}

\subsubsection{Dense Ellipsoid: }
\label{exp:ellipsoid}
This simulation comprises 1,920 agents. To reach their goals, an
ellipsoid-shaped group of agents (Fig.~\ref{single_boat}), with an
initial separation distance of 3.3, must locomote through a larger,
rectangular group of agents, with a separation distance of 3.0.
Throughout the entire simulation, the small group retains its shape
and successfully passes the larger group.

\subsubsection{Proximal Behavior, Avoidance Model: }
\label{exp:proximal_sliding}
Two groups of 50 agents start in tightly packed formations, and must
pass each other in a narrow hallway with limited collision avoidance
space (Fig.~\ref{double_proxemic}). This benchmark demonstrates that
our novel avoidance model creates proxemic behavior in agent groups
\cite{he:2016:group}.

\subsubsection{Proximal Behavior, Long Range Collision: }
\label{exp:proximal_long}
Here, we used the same setting as \ref{exp:proximal_sliding}. We
observed lane formation and splitting of the original group.

\subsubsection{Target Locomotion, Long Range Collision: }
\label{exp:sitting}
192 agents start in a uniform random grid setting at a separation
distance of 5.5. The locomotion targets are in a similar, but in a
translated grid pattern, randomly perturbed with additive uniformly
distributed random noise. The objective of this benchmark was to show
that agents are able to reach their respective goal with minimal
interference.

\subsubsection{Bottleneck: }
\label{exp:bottle}
We demonstrated our method on a bottleneck scenario with varying
number of agents. Agents must pass through a narrow corridor to reach
their goal (Fig.~\ref{arching}). In this scenario, we observed jamming
and arching near the corridor's entrance, as well as the formation of
pockets, a phenomena observed in realistic crowds, which was also
reported in \cite{golas:2014:cmc,guy2010ple}.

\subsection{Comparison}

The method described in \cite{karamouzas:2014:powerlawPRL} is
considered the state-of-the-art model for explicit force-based
modeling of pedestrian behavior, and it has been validated against
human behavior. We implemented this method based on code obtained from
the authors. For the comparison, we chose the same parameter settings
and time-step as in our method (Section~\ref{sec:params}). Using 1,344
agents, we preformed experiments in the two following settings
(Fig.~\ref{single_force_powerlaw}):

\begin{figure}
\includegraphics[width=\columnwidth]{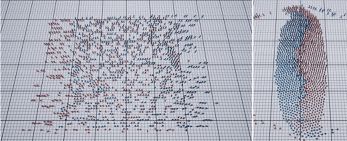}
\caption{Explicit force-based power law
\cite{karamouzas:2014:powerlawPRL}. Left: In a sparse setting, the
agents successfully avoid collisions. Right: In a dense setting, the
agents collide, overlap, and are not able to pass smoothly.
\label{single_force_powerlaw}}
\end{figure}

\subsubsection{Crowd Passing (Sparse): } For the sparse setting, we used a
separating distance of approximately 4.5 between agents. Agents
preformed well and avoided collisions, managing to pass with minimal
interference to the opposing group. Lane patterns emerged.

\subsubsection{Crowd Passing (Dense): } In the dense setting, we used a
separation distance of approximately 3.3. In this setting, the agents
were not able to maintain their trajectory or avoid collisions with
the opposing group. Some of these collisions were not resolved,
leading to an unrealistic state for almost half of the simulation. Our
supplemental video offers additional details.

\bigskip

From the above experiments, we noticed that the power law method does
not provide a collision-free model for dense crowds. Nevertheless,
careful parameter tuning or increasingly small time steps may help,
albeit at the expense of efficiency and ease of use.

\section{Discussion}
\label{sec:discuss}

In this paper, we adapted Position-Based Dynamics (PBD) as an
alternative discrete algorithm for simulating multi-agent dynamics.
Our machinery demonstrated interesting group interactions, such as
groups passing each other seamlessly, as well as the formation of
traffic lanes and subgroups with minimal interference. We demonstrated
our novel PBD method on groups of agents of various sizes, arranged in
varying densities, using different mixtures of PBD constraints. We
presented novel long range collision constraints with adaptive
stiffness, which serve as a realistic preconditioner for the actual
collision from frictional contact, with a sufficient stiffness that
enforces non penetration. Our solution is flexible and produces
interesting patterns and emergent behavior. Compared to existing
methods, the advantages of PBD are large time steps, guaranteed
stability, and ease of control. In addition, our approach allows
simple integration into a preexisting PBD framework. By adding new
constraints, our robust, parallel framework can easily incorporate
more complex crowd behaviors with minimal run time cost.

Nonetheless, our approach has some limitations. We do not pretend to
simulate real pedestrians
(cf.~\cite{Shao:2007,Yu:2007:DNF:1272690.1272707}). Designing metrics
to evaluate such realism is a problem in and of itself, and it is
outside the scope of our present work, but we will investigate this
topic in future work, including further quantitative analysis of
time-to-collision and other anticipatory position analysis. Even
though PBD is a simple and stable framework, it requires a certain
amount of parameter tuning. We also plan to explore other constraints,
such as clamping the magnitude of turning and backwards motion of
agents. We believe that such a constraint will lead to more realistic
results. Finally, experimenting with other online locomotion synthesis
methods such as motion fields \cite{lee2010motion} can lead to more
interesting agent interactions.

\begin{acks}
We thank Rahul Narain for valuable discussions. Jiang gratefully
acknowledges a GPU donation from NVIDIA Corporation and a gift from
Awowd, Inc.
\end{acks}

\newpage
\bibliographystyle{ACM-Reference-Format}
\bibliography{references} 

\end{document}